\documentclass[conference]{IEEEtran}
\IEEEoverridecommandlockouts
\usepackage{cite}
\usepackage{amsmath,amssymb,amsfonts}
\usepackage{algorithmic}
\usepackage{graphicx}
\usepackage{glossaries}
\usepackage{subcaption}
\usepackage{booktabs}

\newacronym{iot}{IoT}{Internet of Things}
\newacronym{nbiot}{NB-IoT}{Narrowband IoT}
\newacronym{ue}{UE}{User Equipment}
\newacronym{at}{AT}{Advanced Technology}
\newacronym{mac}{MAC}{Media Access Control}
\newacronym{rlc}{RLC}{Radio-Link Communication}
\newacronym{lte}{LTE}{Long-Term Evolution}
\newacronym{mtc}{MTC}{Machine-Type Communication}
\newacronym{ltem}{LTE-M}{LTE for Machine Type Communication}
\newacronym{3gpp}{3GPP}{3rd Generation Partnership Project}
\newacronym{etsi}{ETSI}{European Telecommunications Standards Institute}
\newacronym{enb}{eNB}{evolved Node-B}
\newacronym{rsrp}{RSRP}{Reference Signal Received Power}
\newacronym{mse}{MSE}{Mean Square Error}
\newacronym{mae}{MAE}{Mean Absolute Error}
\newacronym{ce}{CE}{Coverage Enhancement}
\newacronym{o2i}{O2I}{Outdoor-To-Indoor}
\newacronym{gps}{GPS}{Global Positioning System}
\newacronym{ols}{OLS}{Ordinary Least Squares}
\newacronym{gnss}{GNSS}{Global Navigation Satellite System}

\def\BibTeX{{\rm B\kern-.05em{\sc i\kern-.025em b}\kern-.08em
    T\kern-.1667em\lower.7ex\hbox{E}\kern-.125emX}}
\begin{document}

\title{Experimental Evaluation of Empirical NB-IoT Propagation Modelling in a Deep-Indoor Scenario}

\author{\IEEEauthorblockN{Jakob Thrane, Krzysztof Mateusz Malarski, Henrik Lehrmann Christiansen and Sarah Ruepp}
\IEEEauthorblockA{\textit{DTU Fotonik} \\
\textit{Technical University of Denmark}\\
Kgs. Lyngby, 2800 DK \\
\{jathr, krmal, hlch, srru\}@fotonik.dtu.dk}

}

\maketitle

\begin{abstract}
Path-loss modelling in deep-indoor scenarios is a difficult task. On one hand, the theoretical formulae solely dependent on transmitter-receiver distance are too simple; on the other hand, discovering all significant factors affecting the loss of signal power in a given situation may often be infeasible. In this paper, we experimentally investigate the influence of deep-indoor features such as indoor depth, indoor distance and distance to the closest tunnel corridor and the effect on received power using NB-IoT. We describe a measurement campaign performed in a system of long underground tunnels, and we analyse linear regression models involving the engineered features. We show that the current empirical models for NB-IoT signal attenuation are inaccurate in a deep-indoor scenario. We observe that 1) indoor distance and penetration depth do not explain the signal attenuation well and increase the error of the prediction by 2-12 dB using existing models, and 2) a promising feature of average distance to the nearest corridor is identified.

\end{abstract}


\section{Introduction}
\begin{figure*} 
    \centering
    \includegraphics[width=\textwidth]{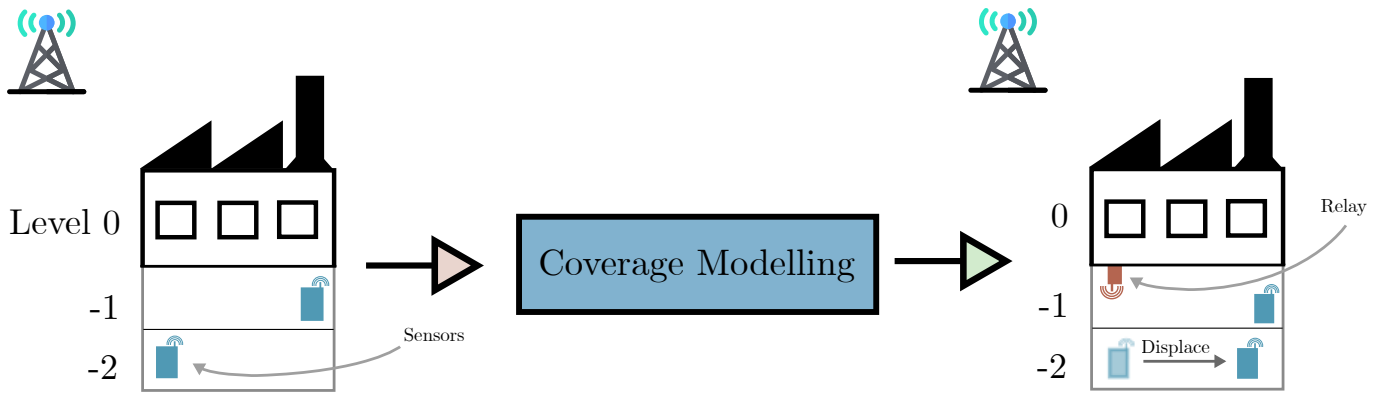}
    \caption{Indoor deployment situations are further complicated by deep-indoor situations such as basements where coverage modelling is difficult and impractical. Current models perform well at indoor deployment situated at level 0 and above but are inaccurate at level -1 and -2. This paper presents measurements conducted at level -1 and -2.}
    \label{fig:modelling}
\end{figure*}
According to IoT analytics, more than a half of the enterprise \gls{iot} projects in 2018 were classified as smart city, connected industry and connected building; in such categories, asset tracking and environment monitoring are prominent use cases \cite{iotan2018}. Applying \gls{iot} to remote monitoring, e.g. smart water metering, the main problem is to ensure reliable connectivity and optimised power consumption of the sensors placed in basements or underground tunnels. The solution must consist of an appropriate hardware design, a suitable communication technology and knowledge of signal behaviour in the deployment area, so that the service provider can guarantee seamless and economically feasible service in the customer's environment.

Cellular \gls{iot} technologies such as \gls{nbiot} and \gls{ltem} are tailored for long-range applications, and they are expected to dominate the market of massive \gls{iot} due to an excellent link budget, long battery lifetime and security and reliability support \cite{gsma1}. Both standards provide advanced power saving mode and discontinuous reception techniques to save energy, and introduce 20dB link budget improvement in comparison to \gls{lte} due to higher power spectral density and message repetition schemes in uplink and downlink \cite{sundberg2018a, chen2017a}. However, \gls{nbiot} additionally enables multiple deployment options (in-band with \gls{lte}, in the \gls{lte} guardband and standalone) and outperforms \gls{ltem} in terms of energy efficiency in low data-rate scenarios and when radio conditions are poor \cite{el2018a}. 

Even with \gls{nbiot} the problem of bad or no coverage in remote, hard-to-reach areas (especially underground) persists. The number of packet repetitions is dictated by the current \gls{ce} level, identified by the network based on the perceived radio conditions \cite{andres-maldonado2017a}. At the same time, the energy usage grows as the number of message repetitions increases \cite{harwahyu2019a}. 
In deep-indoor situations, high signal attenuation causes \gls{nbiot} operation on \gls{ce} levels corresponding to the biggest number of repetitions (up to 128 in uplink), leading to increased power consumption. Thus, understanding signal propagation and attenuation in underground environments is essential in the process of optimal sensor placement and connectivity and throughput modelling.


\gls{3gpp} and \gls{etsi} derived theoretical path-loss models covering outdoor-to-outdoor, outdoor-to-indoor and indoor-to-indoor scenarios \cite{3GPP_38901, Itu-r}. However, the assumptions regarding deep-indoor path-loss are oversimplified for some underground scenarios (see Fig. \ref{fig:modelling}). Specifically, the fact that the attenuation of the signal in the aforementioned theoretical models depends solely on the distance between the transmitter and the receiver may lead to rough conclusions not reflecting other environmental factors. For that reason, investigating new features related to the communication scenario appeals promising. 

A comprehensive survey on radio propagation modelling in deep-indoor propagation situations and tunnel systems can be found in \cite{Hrovat2014}. The authors discuss several modelling techniques for radio propagation in tunnels hereof the use of ray-tracing and empirical models. The theoretical analysis shows that tunnel geometry have an important impact on the attenuation rate of the received power which is not taken into account by empirical models thus leading to inaccurate predictions.

In this work, we present our efforts toward better understanding of deep-indoor path loss of \gls{nbiot}. We conducted a measurement campaign and collected radio signal strength samples from a \gls{nbiot} device. We observed that the received power does not decrease with the transmitter receiver separation distance. This led us to derive more parameters - indoor depth, indoor distance and average distance to the closest corridor, and to study their significance to signal attenuation. The main contributions of the paper can be summarised as follows:

\begin{itemize}
    \item We present a unique measurement campaign, performed in the underground tunnels and basements of the Technical University of Denmark, spanning the entirety of campus.
    \item We formulate the following features: indoor depth, indoor distance and average distance to the closest corridor.
    \item We discuss the significance of the considered parameters in modelling the path loss of \gls{nbiot} in deep-indoor environments and open issues concerning underground deployments and coverage studies.
\end{itemize}

The remainder of this paper is organised as follows. We introduce the available path-loss models and motivate the study in Section \ref{sec:methodology}. The formulation of the features is explained in Section \ref{sec:new_feats}. The description of our measurement campaign and the primary data analysis are included in Section \ref{sec:measurements}. Section \ref{sec:analysis} contains the statistical analysis of the engineered features, and further discussion on general issues is included in Section \ref{sec:discussion}. We conclude the study in Section \ref{sec:conclusions}.



\section{Methodology}
\label{sec:methodology}

The ultimate goal of coverage modelling is to obtain realistic signal propagation behaviour, the analysis of which constitutes to more optimised real-life deployment. Apart from reflecting the field measurements faithfully, the model ought also to be generalised, in other words, applicable to more scenarios than the one accompanying model formulation. Deterministic models (e.g. Ray Tracing) take into account detailed profile of the environment, thus produce reliable predictions. However, they are computationally complex and biased towards the particular scenario. On the other hand, statistical models are simpler and more general, as they consider only limited set of variables explaining the signal attenuation, and they do not take into account the particularities of any specific environment; yet, the accuracy of the statistical models depends on the amount of available measurement data used for model derivation.

\subsection{Outdoor-to-Indoor path-loss}\label{subsec:o2i_3gpp}
The approach for \gls{o2i} path-loss modelling is described in \cite{3GPP_38901} and utilise a sequence of necessary steps. The path-loss is decomposed into several terms as given below:

\begin{equation}\label{eq:PL_o2i}
    \text{PL}_{o2i} =  \text{PL}_b +  \text{PL}_{tw} +  \text{PL}_{in} + \mathcal{N}(0,\sigma_p^2)
\end{equation}

Where, $\text{PL}_b$ is the \emph{basic} outdoor path-loss, $\text{PL}_{tw}$ are losses associated with building penetration loss (constant and frequency dependent),  $\mathcal{N}(0,\sigma_p^2)$ is a log-normal distribution with local variability $\sigma_p$ and  $\text{PL}_{in}$ are losses dependent on the depth inside the building. However, the model is only defined for \gls{o2i} scenarios with regular buildings and does not consider the indoor depth. The losses associated with the indoor distances are given as follows:

\begin{equation}\label{eq:PL_in}
     \text{PL}_{in} = 0.5 \cdot d_{in,2d}
\end{equation}

Where $d_{in,2d}$ is the distance indoor, e.g. the distance to the outer most wall closest to the transmitter. In a basement scenario this parameter is unspecified. The primary contribution of this paper is evaluating such indoor depth parameters for path-loss modelling.

\subsection{Statistical analysis}

Since the deep-indoor loss component of the official path-loss model is linear, we studied the relevance of the engineered features by means of linear regression; we applied \gls{ols} technique \cite{seabold2010statsmodels} and compared determination coefficient $R^{2}$, Log-likelihood and Residual \gls{mse} statistics.

\section{Feature engineering}
\label{sec:new_feats}

Received power of the signal decreases with increased distance as denoted by basic path-loss models, however, in the outdoor-to-deep-indoor scenario, penetrating multiple media (air, outdoor obstacles, ground, tunnel walls) makes the power-distance relationship more complex. In practice, it is difficult to know the exact characteristics of all the materials through which the wave would penetrate, or even the kind of the materials from which e.g. the underground constructions are made. Furthermore, engineering features for path-loss estimation that is capable of explaining such complex interactions is problematic due to inaccuracy of obtaining indoor positions. In this paper we obtain the indoor positions and the features using a massive and high resolution LIDAR dataset of the entire tunnel system. 

\begin{figure} [h]
    \centering
    \includegraphics[scale=0.3]{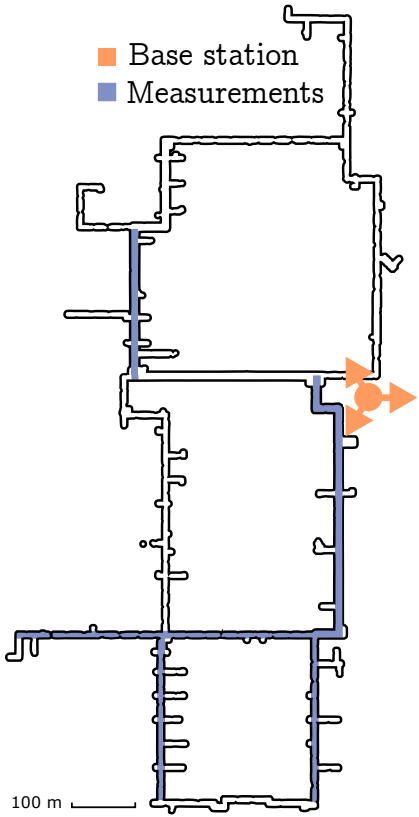}
    \caption{Layout of tunnel system where measurements were conducted. The tunnel system is considered between level -1 and -2. The base station is placed at $30$ m above top ground level. }
    \label{fig:setup}
\end{figure}

\subsection{Indoor positioning}\label{subsec:positioning}
Indoor positioning is a non-trivial task as common ways of obtaining positions (e.g. using \gls{gnss} solutions) are not possible in indoor and deep-indoor situations. Several techniques for obtaining indoor positions based on radio waves are documented in literature \cite{Nuaimi2011} but require existing infrastructure and complex fingerprinting implementations. In this work we have access to a high resolution LIDAR dataset of the measured area (see Fig. \ref{fig:setup}). The entirety of the tunnel area is sampled in $(x,y,z)$ coordinate points with a resolution of $<1$ \; cm. In order to utilise such a massive dataset we used the following procedure for identifying the indoor positions. 1) each independent measurement study was composed of a starting position and an end position; 2) the start and end positions were identifying in the LIDAR dataset (point cloud) and thus the \gls{gps} positions were extracted; 3) Using the known amount of measurements of the given corridor, in combination with the start and end position, allowed for an interpolation between the equidistant measurement points, thus giving an indoor position (with altitude information) per measurement point.

\subsection{Defining the features}
\begin{figure}
    \centering
    \includegraphics{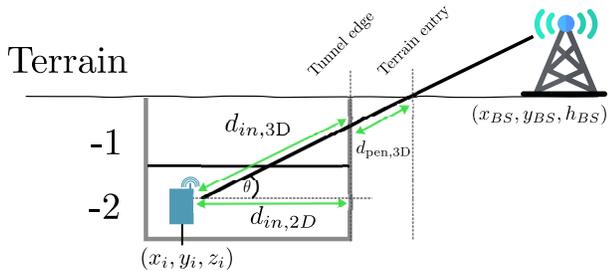}
    \caption{Distance indoor is computed in 2D and 3D using LIDAR information of the tunnel system. The tunnel dimensions and the terrain entry point can be determined using the point cloud and the angles (azimuth and elevation) deduced from the measurement positions}
    \label{fig:inside_distance}
\end{figure}

Having access to a LIDAR dataset with high resolution enabled accurate feature engineering in 3D along with accurate indoor positioning. The configuration of the \gls{nbiot} transmitter is known, including the altitude information, \gls{gps} position and transmitter specific parameters as seen in Table. \ref{tab:params}. The accurate 3D position of the measurements in combination with the details of the transmitter allows for computation of azimuth and elevation angles for each measurement position. Additionally, the LIDAR dataset enabled more advanced features to be engineered which is the primary contribution of this paper. Using the point cloud of the LIDAR dataset, the tunnel dimensions was quantified using 3D trigonometry. This furthermore enabled the engineering of complex features such as the indoor distance ($d_{in}$), the penetration distance ($d_{pen}$), and the average distance to the nearest corridor ($d_{cor,avg}$). Both $d_{in}$ and $d_{pen}$ is computed in a "as-the-crow-flies" path towards the \gls{enb} as illustrated in Fig. \ref{fig:inside_distance} using the elevation and azimuth angles relative to the measurement position. $d_{cor,avg}$ is computed by identifying the corridors crossing the main tunnel of the equidistant measurements using the LIDAR data. Using trigonometry, the average distance to the nearest corridor can be derived. All of the features are derived in 2D and 3D space, i.e. with and without the use of the elevation angle.


\begin{table}[b]
\centering
\begin{tabular}{@{}l|l@{}}
\toprule
\# of measurement points       & $895$             \\ \midrule
TSMW/UE measurements per point & $1e6/10$          \\ \midrule
Operating frequency            & $820.5$ MHz       \\ \midrule
Bandwidth                      & $180$ kHz         \\ \midrule
Noise figure (TX/RX)           & $5$ dB/$3$ dB     \\ \midrule
TX power                       & $46$ dBm          \\ \midrule
Receiver antenna position      & Vertical          \\ \midrule
TX/RX antenna gain             & $5$ dBi/$5.8$ dBi \\ \bottomrule
\end{tabular}
\caption{Experiment parameters}\label{tab:params}
\end{table}

\begin{figure*}[ht]
    \centering
    \begin{subfigure}[b]{0.42\textwidth}
        \centering
        \includegraphics[width=\textwidth]{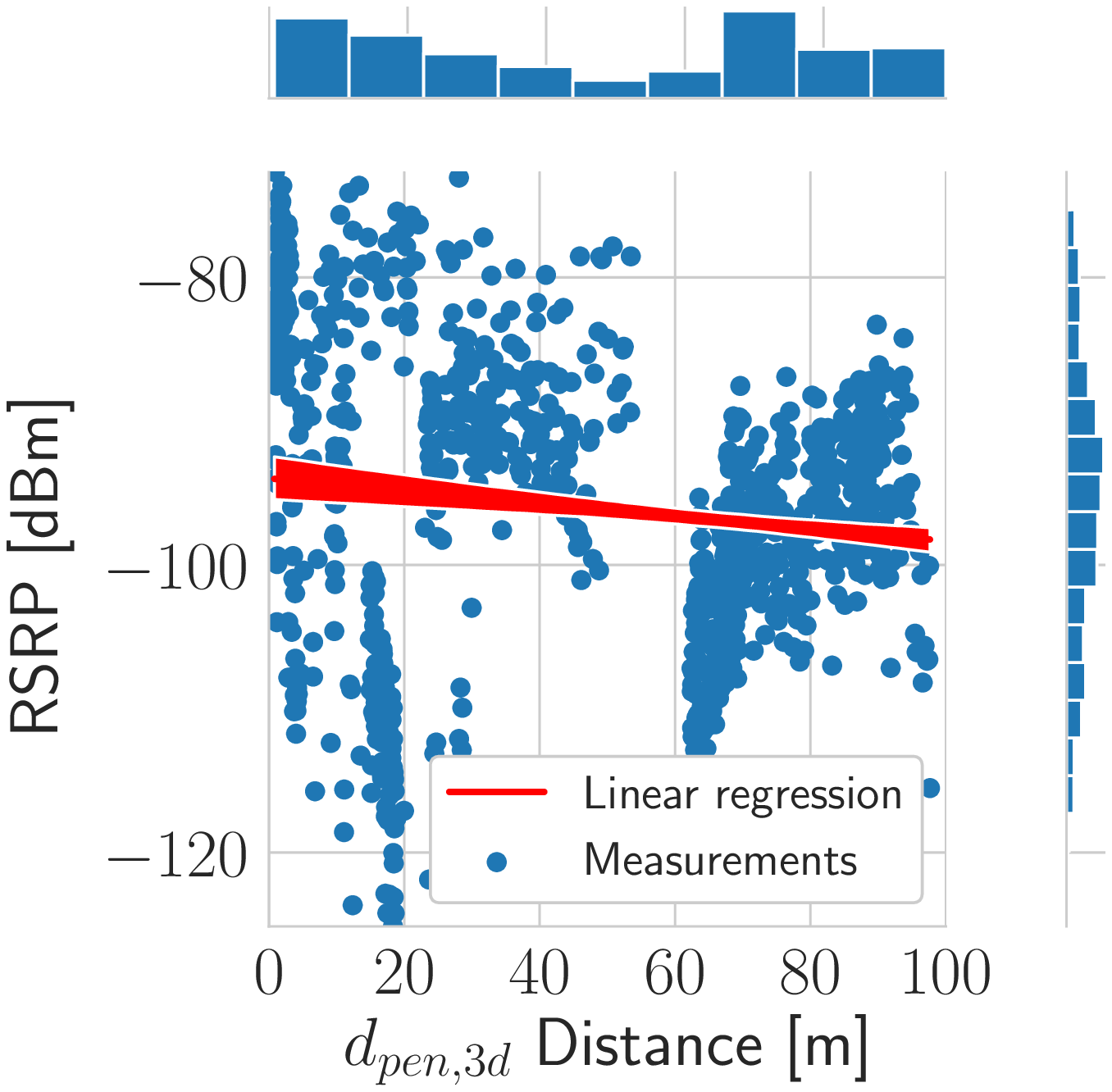}
        \caption{RSRP vs. $d_{pen,3d}$}
    \end{subfigure}%
    ~
    \begin{subfigure}[b]{0.42\textwidth}
        \centering
        \includegraphics[width=\textwidth]{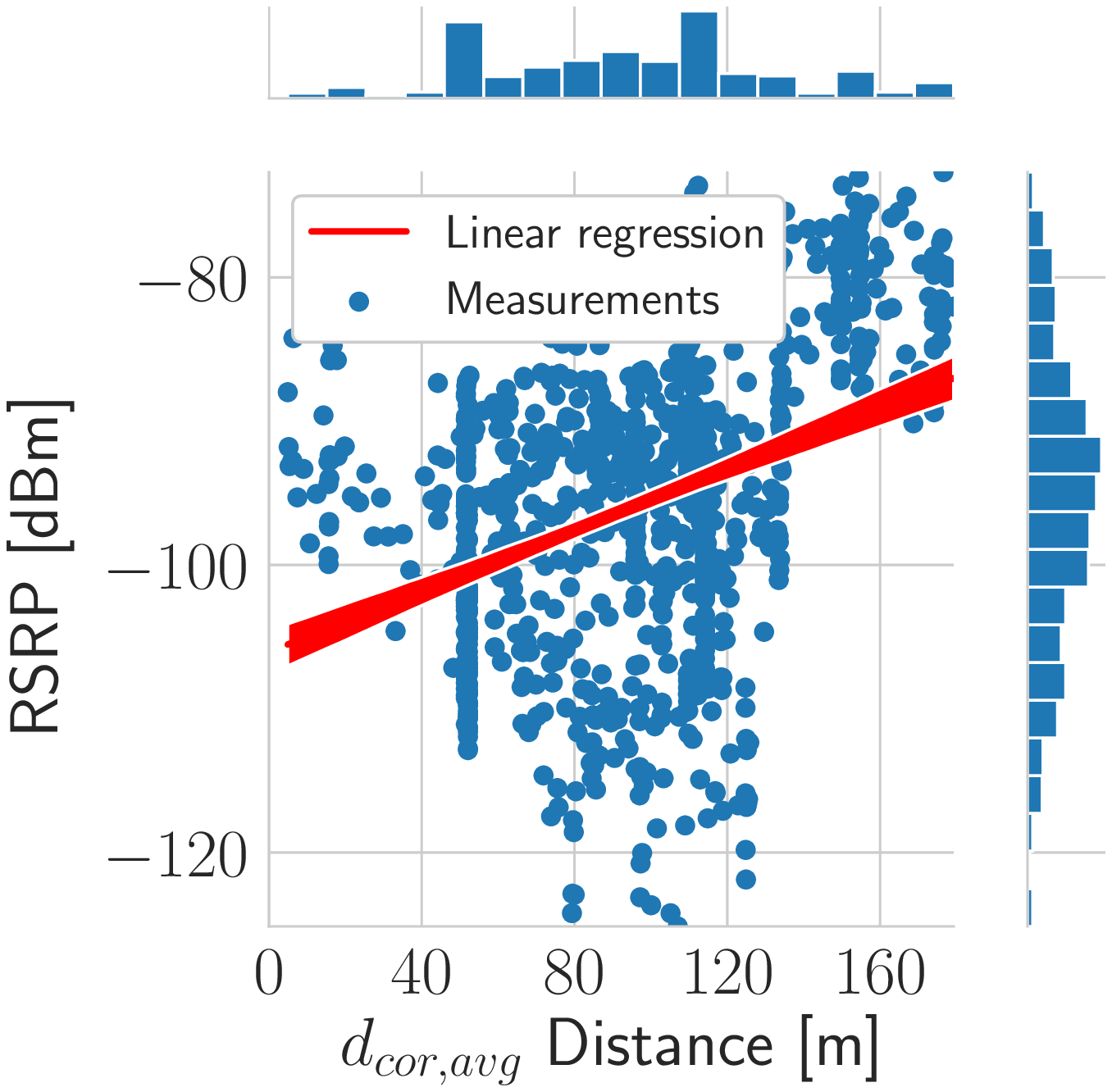}
        \caption{RSRP vs. $d_{cor, avg}$}
        \label{subfig:dcor_rsrp}
    \end{subfigure}%
    \caption{The relationship between RSRP and the engineered features.}
    \label{fig:features}
\end{figure*}

The distributions of the features and their relationship with the \gls{rsrp} of the \gls{nbiot} signal are presented in Fig. \ref{fig:features}. There is a slight and non-linear tendency that the signal attenuates with the growth of indoor depth, but in the case of indoor distance the trend is opposite. A more distinct relationship between the \gls{rsrp} and $d_{cor, avg}$ is visible in Fig. \ref{subfig:dcor_rsrp}.




\section{Measurement campaign}
\label{sec:measurements}

We collected \gls{nbiot} \gls{rsrp} measurements and other \gls{ue} radio statistics from 895 measurement positions within the DTU tunnel system in Lyngby Campus. The area covered by the measurements can be seen in Fig. \ref{fig:setup}. Each of the measured corridors was divided into a set of equidistant locations (1 or 2 metres distance between two measurement positions), and the measuring equipment captured the samples at these distinct locations only, i.e. the measurements were taken stationary.

The setup consisted of a Rohde\&Schwartz TSMW network tester \cite{Manual2017}, u-blox SODAQ SARA N211 \gls{nbiot} device \cite{sodaq}, a laptop and a gel rechargeable battery. The antennae of TMSW and the \gls{ue} were fixed vertically on the trolley. At each of the measurement positions, a low-pass filter around the operating frequency was used to capture $1e6$ NB-IoT IQ samples. Parallel to this 10 measurements of \gls{ue} statistics was obtained using the \gls{nbiot} device. The mean of the measurements was taken to remove any fast-fading impairments. 

\begin{figure*}
    \centering
    \begin{subfigure}[b]{0.4\textwidth}
        \centering
        \includegraphics[width=1\textwidth]{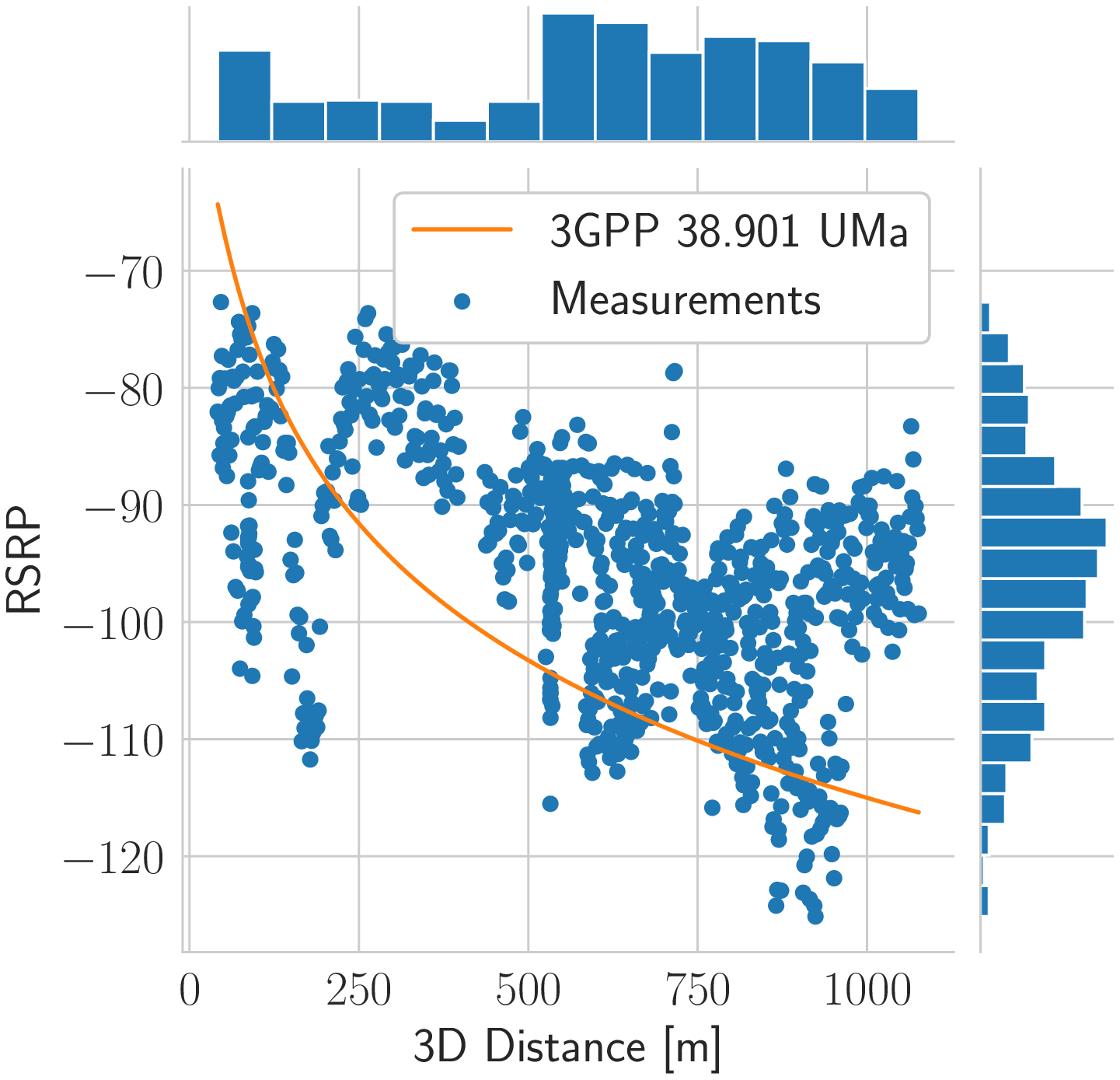}
        \caption{RSRP wrt. 3D distance vs. 3GPP path-loss model}
        \label{subfig:3gpp}
    \end{subfigure}%
    ~ 
    \begin{subfigure}[b]{0.4\textwidth}
        \centering
        \includegraphics[width=1\textwidth]{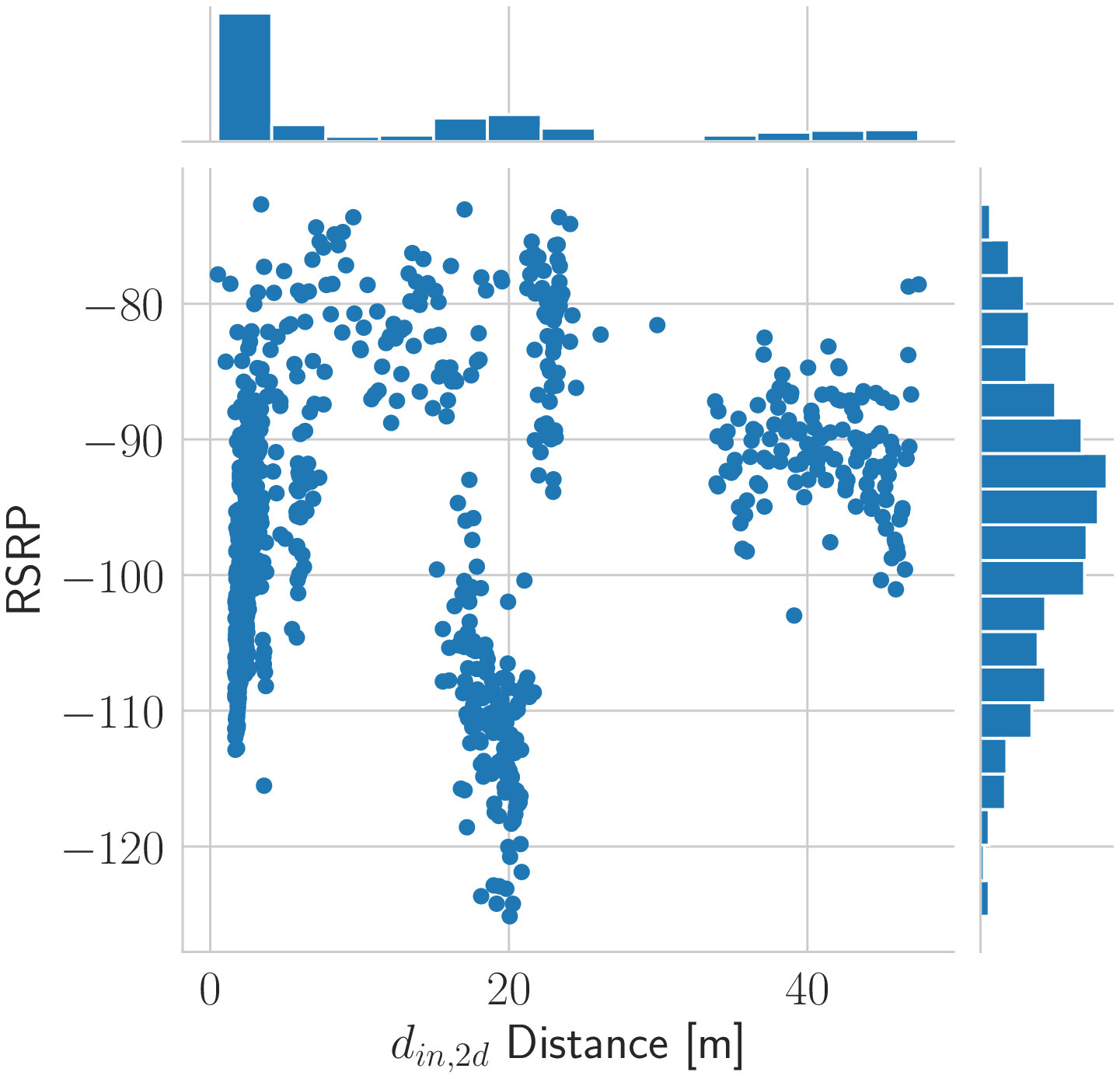}
        \caption{RSRP vs. $d_{in,2d}$}
        \label{subfig:din_rsrp}
    \end{subfigure}
    \caption{Comparison between the observed power-distance relationship and the theoretical 3GPP model.}
    \label{fig:3gpp_comparison}
\end{figure*}

\subsection{Visualisation}

A scatterplot with linear regression fit and histograms in Fig. \ref{fig:3gpp_comparison} visualises the nature and mutual relation of \gls{rsrp} and 3D distances between the \gls{ue} and the \gls{enb}. It is possible to notice that \gls{rsrp} does not depend linearly on the distance, as their distributions are clearly different; one may observe that the line representing the \gls{3gpp} model fits the experimental observations poorly. This agrees with the findings of our previous study, described in \cite{malarski2019a}, but now proven over larger measurement area. Interestingly, the behaviour of \gls{rsrp} with respect to the indoor distance is not linear either. Therefore, we believe that other features are needed to fully explain the complex behaviour of \gls{nbiot} signal attenuation underground.

\section{Results}
\label{sec:analysis}
\subsection{Linear regression}
\label{subsec:reg}

Table \ref{tab:reg} presents basic statistic of linear regression fit of the investigated features on \gls{rsrp}. Additionally, we added a regression model employing azimuth angle $\phi$ and elevation angle $\theta$. These parameters are not considered useful in path-loss modelling, but were included in the statistical analysis as a source of reference to better evaluate the indoor features. 

Model M1 with 3D distance exhibits the lowest \gls{mse} (74.973) and the highest $R^2$ coefficient (0.285). On the other hand, M4 combining $d_{pen, 3D}$ and $d_{in, 2D}$ yields 0.005 $R^2$ and 104.335 \gls{mse}. Noteworthy, \gls{mse} of $d_{cor, avg}$ model is lower than in $d_{pen, 3D}$ and $d_{in, 2D}$ models by 13.934 and 12.812, respectively, and $R^2$ is higher by 0.131 and 0.122, respectively.


\begin{table}[b]
\caption{Summary of linear regression statistics}
\centering
\begin{tabular}{|l|l|l|l|l|}
\hline
ID & Regressors                                                                               & R\textasciicircum{}2 & Log-likelihood & Residual MSE \\ \hline
M1       & 3D distance                                                                              & 0.285                & -3200.9        & 74.973       \\ \hline
M2       & $d_{in,2D}$                                                                       & 0.026                & -3389.1        & 102.098      \\ \hline
M3       & $d_{pen,3D}$                                                                             & 0.017                & -3343.1        & 103.022      \\ \hline
M4       & \begin{tabular}[c]{@{}l@{}}$d_{pen,3D}$ + $d_{in,2D}$\end{tabular}                 & 0.005                & -3348.8        & 104.335      \\ \hline

M5       & $d_{cor,avg}$                 & 0.148                & -3279        & 89.286      \\ \hline
M6       & \begin{tabular}[c]{@{}l@{}}$d_{pen,3D}$ + $d_{in,2D}$\\ + $d_{cor,avg}$\end{tabular} & 0.150                & -3278        & 89.276       \\ \hline
M7       & \begin{tabular}[c]{@{}l@{}}$\phi$ + $\theta$\end{tabular}                 & 0.173                & -3265.7        & 86.763      \\ \hline
\end{tabular}
\label{tab:reg}
\end{table}

\subsection{Indoor distance features}
\begin{figure}
    \centering
    \includegraphics[width=0.5\textwidth]{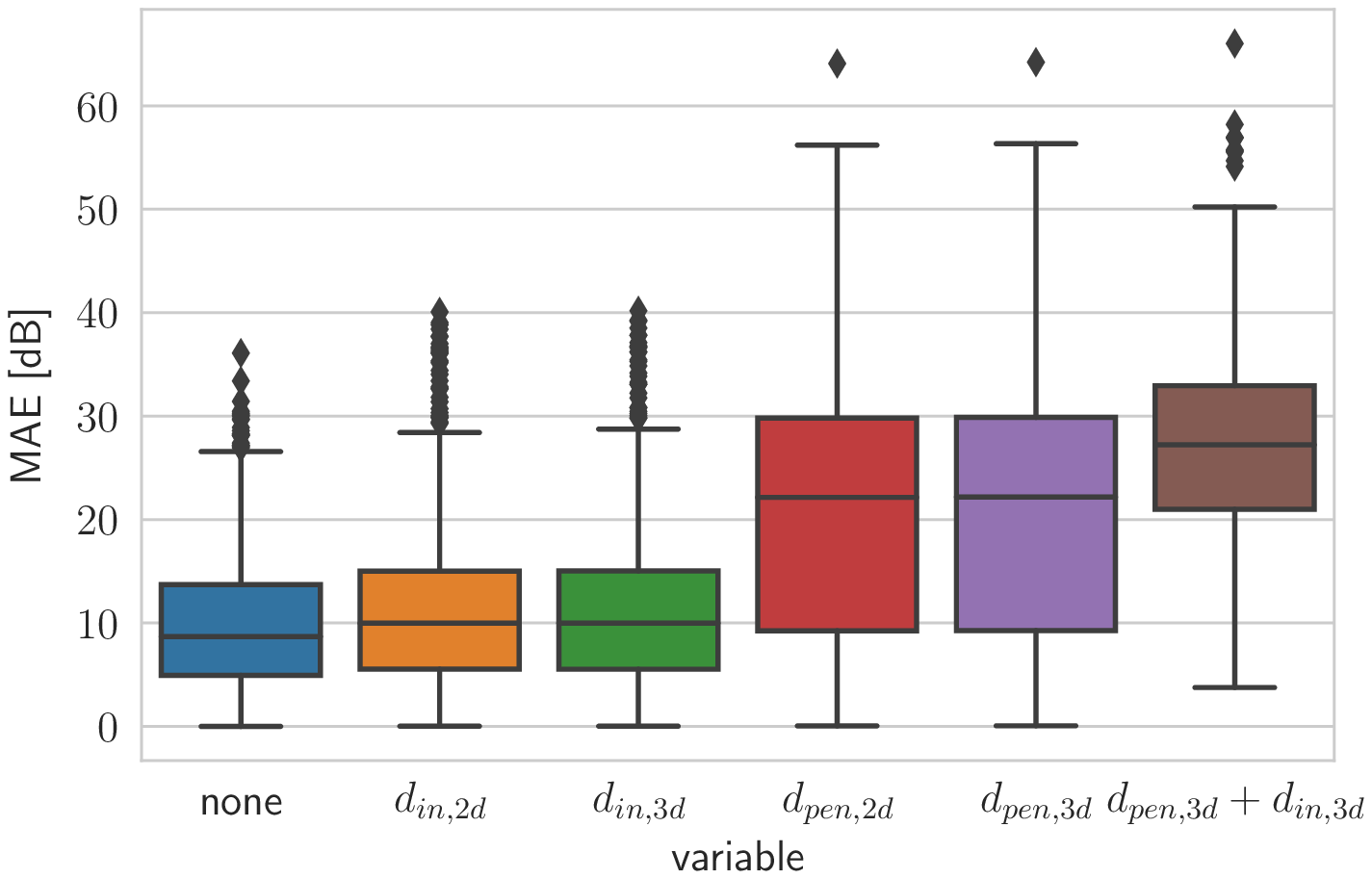}
    \caption{\gls{mae} of utilising different indoor distance and penetration depth features in Eq. (\ref{eq:PL_in})}
    \label{fig:mae_boxplot}
\end{figure}

The \gls{o2i} modelling principles as detailed in Section \ref{subsec:o2i_3gpp} is undefined for basement scenarios. Thus, a prediction comparison utilising the penetration distance, and the indoor distances as \emph{indoor distances} in accordance with Eq. (\ref{eq:PL_in}) are shown in Fig. \ref{fig:mae_boxplot}. The \emph{none} case defines the use of path loss principles for \gls{o2i} scenarios using Eq. (\ref{eq:PL_o2i}) but without the $\text{PL}_{in}$ term. The remainder of the plot shows the \gls{mae} prediction errors as a function of different indoor distance parameters. It is found that utilising any of the indoor distance metrics in this particular basement scenario increases the prediction error by $\approx 2$ dB to $\approx 12$ dB.


\section{Discussion}
\label{sec:discussion}

Based on Tab. \ref{tab:reg} it can be observed that none of the parameters nor combinations thereof perform better than 3D distance between the \gls{ue} and the \gls{enb} (model M1). Moreover, $d_{in, 2D}$ and $d_{pen, 2D}$ explain only marginal share of the  \gls{rsrp} variance and exhibit the highest \gls{mse} (M2-M4). On the other hand, model M5 involving $d_{avg, cor}$ feature, as well as model M7 consisting of $\phi$ and $\theta$ angles yield significantly better results. This indicates that indoor distance and indoor depth are not useful in deep-indoor path-loss modelling. Instead, the features related to the underground corridors (e.g. $d_{avg, cor}$) and/or other geographical phenomena represented here by model M7 should be considered.

\subsection{Application considerations}

In the former part of this paper we evaluated the engineered features in terms of statistical metrics, however, in order to apply the features in real-life \gls{nbiot} scenarios, such as smart metering or underground monitoring, the following aspects need to be considered.

\subsubsection{Indoor positioning problem}

Computing indoor depth and indoor distance can only be done knowing the precise location of the \gls{ue} and the \gls{enb}. In our case, the availability of LIDAR point cloud was essential, as it enabled to deduce the measurement points with ca. 50 cm precision. Albeit, one certainly cannot rely on such data in an arbitrary deep-indoor area, and the fact that global localisation systems, such as GPS or GNSS are unreachable underground means that knowing where the device resides can be difficult. 

\subsubsection{Significance of other environmental features}

Even though the results presented in Section \ref{subsec:reg} exhibit some correlation between the indoor parameters and the \gls{rsrp}, a stronger relationship comes from $\phi$ and $\theta$ angles, which point at other features describing the measurement area and not being directly associated with indoor penetration. It is enough to mention the following: the footprint of the buildings, the size and structure of tunnel corridors and ventilation ducts or thickness of the entry doors. Moreover, the presence of machines, pipes or solid structures inside the considered underground area may also influence the coverage situation significantly. Last, but not least, one must not forget the impact of above-ground buildings and other obstacles on power losses, even deep-indoor. Evaluating the aforementioned features is out of scope of this paper.

\subsubsection{Feasibility}

Examining Table \ref{tab:reg} one may put in question the sense of finding indoor depth/distance features to apply them as signal power regressors; the \gls{mse} is considerably higher than in the case of total 3D distance, which is easier to compute knowing the locations of the transmitter and the receiver. Furthermore, the features alone explain less than 3\% variance, which may lead to a fundamental question: should one rely on indoor depth and indoor distance in coverage prediction, or would it be more convenient to conduct trial-and-error tests instead?

As a matter of fact, not only path-loss and coverage modelling plays an important role in deep-indoor IoT service deployment planning. For instance, it is essential to provide all the devices for energy, either in a form of batteries (then device accessibility and possibility of battery replace is a key) or, possibly, with the use of locally deployed electrical installation. Moreover, in industrial scenarios, the presence of other devices or machinery might potentially cause interference.

Since we observed a significant share of variance explained by the elevation and bearing angles, we conclude that in underground scenarios the complex behaviour of signal attenuation is primarily caused by geographical parameters of the environment not explained by the features presented in this work. Discovering, engineering and analysing other parameters has been left for future work.

\section{Conclusions}
\label{sec:conclusions}

In this paper we present a measurement campaign conducted in an underground tunnel system. With the aid of LIDAR point cloud data of the tunnels, we deduced the precise locations of the measurement points and, besides 3D distance, we derived 3 more parameters: indoor depth, indoor distance and average distance to the closest corridor. A basic statistical analysis of the linear regression models revealed that indoor distance features (indoor depth and indoor distance) are not explanatory and alone cannot constitute a good approximator for margin budgets in deep indoor situations. Additionally, it is shown that current empirical models offer poor prediction performance using such indoor distance metrics. Instead, features unrelated to indoor distance (such as the average distance to the closest corridor) represent stronger correlation to the signal attenuation and should be further investigated for use in empirical path loss models.


\section*{Acknowledgement}
This work was partially supported by Innovation Fund Denmark through the Eureka Turbo project IoT Watch4Life.

\bibliographystyle{./IEEEtran}
\bibliography{ms.bib}

\begin{thebibliography}{10}
\providecommand{\url}[1]{#1}
\csname url@samestyle\endcsname
\providecommand{\newblock}{\relax}
\providecommand{\bibinfo}[2]{#2}
\providecommand{\BIBentrySTDinterwordspacing}{\spaceskip=0pt\relax}
\providecommand{\BIBentryALTinterwordstretchfactor}{4}
\providecommand{\BIBentryALTinterwordspacing}{\spaceskip=\fontdimen2\font plus
\BIBentryALTinterwordstretchfactor\fontdimen3\font minus
  \fontdimen4\font\relax}
\providecommand{\BIBforeignlanguage}[2]{{%
\expandafter\ifx\csname l@#1\endcsname\relax
\typeout{** WARNING: IEEEtran.bst: No hyphenation pattern has been}%
\typeout{** loaded for the language `#1'. Using the pattern for}%
\typeout{** the default language instead.}%
\else
\language=\csname l@#1\endcsname
\fi
#2}}
\providecommand{\BIBdecl}{\relax}
\BIBdecl

\bibitem{iotan2018}
\BIBentryALTinterwordspacing
IoT-analytics, ``The top 10 iot segments in 2018 – based on 1,600 real iot
  projects,'' accessed: 30-03-2020. [Online]. Available:
  \url{https://iot-analytics.com/top-10-iot-segments-2018-real-iot-projects/}
\BIBentrySTDinterwordspacing

\bibitem{gsma1}
\BIBentryALTinterwordspacing
GSMA, ``Mobile iot. market opportunity for low power wide area networks.''
  accessed: 23-03-2020. [Online]. Available:
  \url{https://www.gsma.com/iot/mobile-iot/}
\BIBentrySTDinterwordspacing

\bibitem{sundberg2018a}
M.~Sundberg, J.~Sachs, J.~Bergman, Y.-P.~E. Wang, and O.~Liberg,
  \emph{\BIBforeignlanguage{eng}{Cellular Internet of Things}}.\hskip 1em plus
  0.5em minus 0.4em\relax Academic Press, 2018.

\bibitem{chen2017a}
M.~Chen, Y.~Miao, Y.~Hao, and K.~Hwang, ``\BIBforeignlanguage{eng}{Narrow band
  internet of things},'' \emph{\BIBforeignlanguage{eng}{Ieee Access}}, vol.~5,
  pp. 8\,038\,776, 20\,557--20\,577, 2017.

\bibitem{el2018a}
M.~El~Soussi, P.~Zand, F.~Pasveer, and G.~Dolmans,
  ``\BIBforeignlanguage{eng}{Evaluating the performance of emtc and nb-iot for
  smart city applications},'' \emph{\BIBforeignlanguage{eng}{Ieee International
  Conference on Communications}}, vol. 2018-, p. 8422799, 2018.

\bibitem{andres-maldonado2017a}
P.~Andres-Maldonado, P.~Ameigeiras, J.~Prados-Garzon, J.~Navarro-Ortiz, and
  J.~M. Lopez-Soler, ``\BIBforeignlanguage{eng}{Narrowband iot data
  transmission procedures for massive machine-type communications},''
  \emph{\BIBforeignlanguage{eng}{Ieee Network}}, vol.~31, no.~6, pp.
  8\,120\,238, 8--15, 2017.

\bibitem{harwahyu2019a}
R.~Harwahyu, R.~G. Cheng, W.~J. Tsai, J.~K. Hwang, and G.~Bianchi,
  ``\BIBforeignlanguage{eng}{Repetitions versus retransmissions: Tradeoff in
  configuring nb-iot random access channels},''
  \emph{\BIBforeignlanguage{eng}{Ieee Internet of Things Journal}}, vol.~6,
  no.~2, pp. 8\,605\,340, 3796--3805, 2019.

\bibitem{3GPP_38901}
3GPP, ``{TR 138 901 - V14.3.0 - 5G; Study on channel model for frequencies from
  0.5 to 100 GHz (3GPP TR 38.901 version 14.3.0 Release 14)},'' 3GPP, Tech.
  Rep., May 2017.

\bibitem{Itu-r}
\BIBentryALTinterwordspacing
Itu-r, ``{Guidelines for evaluation of radio interface technologies for
  IMT-2020 M Series Mobile, radiodetermination, amateur and related satellite
  services},'' ITU-R, Tech. Rep. M.2412-0, October 2017. [Online]. Available:
  \url{http://www.itu.int/ITU-R/go/patents/en}
\BIBentrySTDinterwordspacing

\bibitem{Hrovat2014}
A.~{Hrovat}, G.~{Kandus}, and T.~{Javornik}, ``A survey of radio propagation
  modeling for tunnels,'' \emph{IEEE Communications Surveys Tutorials},
  vol.~16, no.~2, pp. 658--669, 2014.

\bibitem{seabold2010statsmodels}
S.~Seabold and J.~Perktold, ``statsmodels: Econometric and statistical modeling
  with python,'' in \emph{9th Python in Science Conference}, 2010.

\bibitem{Nuaimi2011}
K.~{Al Nuaimi} and H.~{Kamel}, ``A survey of indoor positioning systems and
  algorithms,'' in \emph{2011 International Conference on Innovations in
  Information Technology}, 2011, pp. 185--190.

\bibitem{Manual2017}
\BIBentryALTinterwordspacing
Rohde\&Schwarz, \emph{{R{\&}S TSMW Universal Radio Network Analyzers User
  Manual}}, 2017.
\BIBentrySTDinterwordspacing

\bibitem{sodaq}
\BIBentryALTinterwordspacing
SODAQ, ``Sodaq sara arduino form factor (aff) n211 including pcb antenna,''
  accessed: 30-03-2020. [Online]. Available:
  \url{https://shop.sodaq.com/sodaq-sara-aff-n211.html}
\BIBentrySTDinterwordspacing

\bibitem{malarski2019a}
K.~M. Malarski, J.~Thrane, M.~G. Bech, K.~Macheta, H.~L. Christiansen, M.~N.
  Petersen, and S.~R. Ruepp, ``\BIBforeignlanguage{eng}{Investigation of deep
  indoor nb-iot propagation attenuation},''
  \emph{\BIBforeignlanguage{eng}{Proceedings of 2019 Ieee 90th Vehicular
  Technology Conference}}, vol. 2019-, p. 8891414, 2019.

\end{thebibliography}

\end{document}